\documentclass[graybox]{svmult}
\usepackage{mathptmx}       
\usepackage{helvet}         
\usepackage{courier}        
\usepackage{makeidx}         
\usepackage{graphicx}        
\usepackage{multicol}        
\usepackage[bottom]{footmisc}
\usepackage{amsmath,amssymb,amsfonts}        
\makeindex             
\usepackage{stmaryrd}
\def\+{{+\!\!\!+}}
\def\pp{\mbox{\tiny${}_{\stackrel\+ =}$}}
\newcommand{\bbD}[1]{\mathbb{D}_{#1}}
\newcommand{\bbDB}[1]{\bar{\mathbb{D}}_{#1}}
\newcommand{\bbX}[1]{\mathbb{X}^{#1}}

\newcommand{\re}[1] {(\ref{#1})}

\begin{document}
\title*{$2d$  Sigma Models and Geometry}
\author{Ulf Lindstr\"om}
\institute{Ulf Lindstr\"om \at METU, Ankara, and Uppsala University email: {ulf.lindstrom@physics.uu.se}}
%
%
\maketitle

\abstract{Supersymmetric nonlinear sigma models have target spaces that carry interesting geometry. The geometry is richer the more supersymmetries the model has. The study of models with two dimensional world sheets is particularly rewarding since they allow for torsionful geometries. In this review I describe and exemplify the relation of $2d$ supersymmetry to Riemannian, complex, bihermitian, $(p,q)$ hermitean, K\"ahler, hyperk\"ahler, generalised geometry and more. \footnote{UUITP-10/22. Prepared for the proceedings of the MATRIX Research Program:
2D Supersymmetric Theories and
Related Topics Melbourne 17 - 28  Jan 2022}
}

\section{Introduction}
\label{sec:1}
Supersymmetric  models are closely associated to complex geometry in several different ways. 
To be able to write some extended models in a manifest form superspaces may be extended with a $CP^1$ at each point thus relating them to twistors \cite{Lindstrom:2009afn}, conformal supergravity can be formulated in terms of local twistors \cite{Howe:2020xrg}, \cite{Howe:2020hxi}, and supersymmetic nonlinear sigma models is often best formulated in terms of complex superfields and typically have complex target space geometries \cite{Zumino:1979et},\cite{Alvarez-Gaume:1981exv}. This last property is what shall concern us here, although we have to mainly restrict to two-dimensional models with two left and two right going supersymmetries i,e,  to $2d$, $(2,2)$ supersymmetry. Fortunately this case is sufficiently rich to warrant independent scrutiny.

The format of the presentation is to first introduce bosonic $2d$ sigma models and then $(1,1)$ and $(2,2)$ supersymmetry. There are essentially three different superfield representations of the $(2,2)$ supersymmetric sigma models and they are presented one by one and their  their target space geometry identified. After the introduction of semichiral superfields and their sigma models,  generalised K\"ahler geometry is defined. This geometry covers all the target space geometries introduced.

\section{Sigma models}
\label{sec:2}
For a review of various applications of sigma models see \cite{Lindstrom:2018aoc}.\\
A  {non-linear sigma model} is a theory of maps from a (super) manifold $\Sigma$ to a target space ${\cal T}$
\begin{align}\nonumber
&X : \Sigma \rightarrow {\cal{T}}\\[1mm]
&X(x)\longmapsto X\in {\cal{T}}~,
\label{eq:01}
\end{align}
with dynamics specified by extremising an action which is, schematically,
\begin{equation}
S=\int_\Sigma dx{\cal{L}}(X)~.
\end{equation}
Its precise form depends on the world volume dimension and the number of supersymmetries. That same  number then constrains the geometry of ${\cal T}$.

\subsection{The bosonic sigma model}
\label{subsec:1}
To be concrete and quickly see the relation to geometry, we consider a bosonic sigma model in $2d$
where the general bosonic sigma model action reads
\begin{equation}
S=\int d^2x \partial_{\+} X^\mu\left( G_{\mu\nu}(X)+ B_{\mu\nu}(X)\right)\partial_{=}X^\nu=:\int d^2x \partial_{\+} X^\mu E_{\mu\nu}\partial_{=}X^\nu~.
\end{equation}
The $2d$ light-cone coordinates are
\begin{equation}
x^\+=x^0+x^1~,~~~~x^==x^0-x^1~.
\end{equation}
The $G$ field is a symmetic tensorfield on ${\cal T}$ which we identify as a {\em metric}.
The $B$ field is a {\em gerbe connection}\footnote{This refers to its global properties, more specifically the behaviour under change of coordinate patches, See e,g,\cite{Hull:2008vw} .} and the action  depends only on its {field strength}
\begin{equation}
H_{\mu\nu\rho}=\partial_{[\rho}B_{\mu\nu]}~.
\end{equation}
as seen directly from the field equations for $X^\mu$ or from the alternative form of the action 
\begin{equation}
S=\int_{\partial V} d^2x \partial_{\+} X^\mu G_{\mu\nu}\partial_{=}X^\nu+{\textstyle \frac 1 3}\int_V d^3x \epsilon^{ijk}\partial_{i}X^{\mu}\partial_{j}X^{\nu}\partial_{k}X^{\rho}H_{\mu\nu\rho}~,
\end{equation}
where the $2d$ space is a boundary of a contractible $3d$ space $V$, and $X$ is extended to live on the $3d$ space. In general there may be different inequivalent extensions which can lead to quantisation conditions on $H$.

The $B$ (or $H$) term is called a {\em Wess-Zumino term}. Since it depends on $B$ only through its field strength $H$, the model is invariant under a $B$ field gauge transformations;
\begin{equation}
B\to B+ d\lambda~, \iff~~B_{\mu\nu}\to B_{\mu\nu}+\partial_{[\mu}\lambda_{\nu]}~.
\end{equation}

The $X^\mu$ field equations that follow from either form of the action read 
\begin{equation}
G_{\mu\nu}\nabla_{\+}^{(-)}\partial_{=}X^\nu=G_{\mu\nu}\nabla_{=}^{(+)}\partial_{\+}X^\nu=0~,
\end{equation}
where the connections now have torsion due to the inclusion of the $B$ field;
\begin{equation}
\Gamma^{(\pm)\mu}_{\sigma\rho}=\Gamma^{(0)\mu}_{\sigma\rho}\pm T_{\sigma\rho}^{~~\mu}~,~~~ T_{\sigma\rho}^{~~\mu}={\textstyle\frac 1 2}H_{\sigma\rho\nu}G^{\nu\mu}~.
\end{equation}

\begin{svgraybox}
We see that the target space of the bosonic sigma model carries
{{\em Riemann geometry with torsion}}. This is also the target space geometry of models with one left and one right supersymmetry, a $(1,1)$ sigma model.
\end{svgraybox}

\subsection{The $(1,1)$ sigma model}
\label{subsec:2}
A general sigma model in $(1,1)$ superspace is
\begin{align}\nonumber
&&S=\int d^2xD_+D_-\Big(D_+\Phi^i(x,\theta)\big(G_{ij}+B_{ij}\big)(\Phi)D_-\Phi^j(x,\theta)\Big)_|\\[1mm]\label{oneoneact}
&&=\int d^2x\partial_{\+}\phi^iE_{ij}(\phi)\partial_{=}\phi^j+....,
\end{align}
where the vertical bar in the first line denotes the $\theta$ independent part. and $\phi=\Phi_|$.
It has $(1,1)$ supersymmetry manifest by construction. The $(1,1)$ algebra is\footnote{In addition to the generators $Q$ of supersymmetry there are spinorial covariant derivatives $D$ such that $\{D,Q\}=0$, The algebra may be defined in terms of either set.}
\begin{equation}
D^2_+=\partial_{\+}~,~~~D^2_-=\partial_{=}
\end{equation}
and the $\Phi^i(x,\theta)$s are $(1,1)$ superfields.
Additional supersymmetries will constrain the geometry further, however. They will have the infinitesimal form \cite{Gates:1984nk}
\begin{equation}
\delta \Phi^i=\epsilon^+J^i_{(+)k}D_+\Phi^k+\epsilon^-J^i_{(-)k}D_-\Phi^k
\end{equation}
The conditions on the $J$s follow from two requirements:
\begin{itemize}
\item Closure of the algebra ~~~~$[\delta_1,\delta_2]\Phi=-2i\epsilon_1\epsilon_2\partial_{}\Phi $
\item Invariance of the action ~~~$\delta S=0$
\end{itemize}

From closure of the algebra it follows that $J^2_{(\pm)}=-1$ and the vanishing of the Nijenhuis tensors:
\begin{equation}
{\cal N}_{J_{(\pm)}}(X,Y)=[X,Y]+{J_{(\pm)}}[J_{(\pm)}X,Y]+{J_{(\pm)}}[X, { J_{(\pm)}}{Y}]- [J_{(\pm)}{X}, {J}_{(\pm)}{Y}]=0~.
\end{equation}
The latter condition ensures integrability of $J_{(\pm)}$ so that there is a global atlas of complex coordinates with holomorphic transition functions.

From invariance of the action it follows that $J_{(\pm)}^tGJ_{(\pm)}=G$, i.e., hermiticity of the metric with respect to both complex structures, and that $\nabla{}^{(\pm)}J_{(\pm)}=0$.

When there are $p-1$ left and $q-1$ right complex structures, we have $(p,q)$ supersymmety. There will then be additional conditions. For example,  $(4,4)$ supersymmetry with vanishing torsion gives hyperk\"ahler geometry
on ${\cal T}$.
 
It is not always possible to find an off-shell manifest superspace formulation of the extended geometry. It {\em is } possible for $(2,2)$ in $2d$ however.

\subsection{ $(2,2)$ sigma models}
\label{subsec:3}
We denote the supersymmetry generators and covariant derivatives by $\mathbb{Q}$ and $\bbD{~}$ respectively. Their anticommutator is required to vanish $\{\mathbb{Q},\bbD{~}\}=0$.
In terms of covariant derivatives 
the $(2,2)$ algebra is,
\begin{equation}
\{\bbD{\pm},\bbDB{\pm}\}=2i\partial_{\pp}~,
\end{equation}
all other (anti)commutators are zero. Since we now have four $\theta$s, the multiplet contained in a general superfield will consist of 16 fields. Such a multiplet is recducible and not suitable for a sigma model description since the lowest bosonic component is not a scalar field. But since the covariant derivatives  anticommute with the supersymmety generators we can use them to impose constraints that will reduce the multiplet.

\subsubsection{Representations of $(2,2)$}
There are three\footnote{Not strictly true. There are also complex linear fields $\Sigma$ satisfying $\bbDB{+} \bbDB{-} \Sigma =0$ and twisted complex linear fields $\tilde \Sigma$ satisfying $\bbDB{+} \bbD{-} \tilde \Sigma =0$ and their complex conjugate, but these are dual to chirals and twisted chirals, respectively. So there is always an equivalrent formulation in terms of the latter fields.} 
types of constrained $(2,2)$ superfields  and correspondingly three different  target space geometries.
\bigskip

\begin{svgraybox}

{\em Chiral superfields } $\Phi$
\begin{equation}
\bbDB{\pm}\Phi=0
\end{equation}
\end{svgraybox}

\begin{svgraybox}

{T\em wisted Chiral superfields} $\chi$
\begin{equation}
\bbDB{+}\chi=0~,~~~\bbD{-}\chi=0
\end{equation}

\end{svgraybox}

\begin{svgraybox}

{\em Left} $\ell$ {\em and Right} $\mathfrak{r}$ {\em Semichiral superfields }
\begin{equation}
\bbDB{+}\ell=0~,~~~\bbDB{-}\mathfrak{r}=0
\end{equation}

\end{svgraybox}
We will introduce the sigma models corresponding to the three kind of fields above and display their target space geometries. We begin with the chiral fields.\\

\centerline{\em Chiral superfields }
~\\
The chiral superfields $\bbDB{\pm}\Phi=0$ are complex and their lowest components will serve as complex coordinates on the target space ${\cal T}$, suggesting that the target space geometry will be complex.
Their component fields are given by
 \begin{align}\nonumber
&\Phi=\phi+\theta^\alpha\psi_\alpha+\theta^\alpha\theta_\alpha{\cal F}~,~~~~~~\alpha={\scriptstyle (+,-)}\\[1mm]
&\phi(x)=\Phi(x,\theta)_|~\\[1mm]\nonumber\label{comps}
&\psi_\alpha(x)=\bbD{\alpha} \Phi(x,\theta)_|~,\\[1mm]
&{\cal F}(x)=\bbD{}^2\Phi(x,\theta)_|~,
\end{align}
where the vertical bar denotes setting  $\theta=0$. This is the preferred way of defining components, whereas the first line represents an expansion in  $\theta$. Such an expansion gets cumbersome when there are more supersymmetries and thus more $\theta$s.

 The most general superspace action for chiral fields reads \cite{Zumino:1979et}
 \begin{equation}
 S=\int d^2x \bbD{}^2\bbDB{}^2K(\Phi,\bar\Phi)~.
 \end{equation}
Pushing in the spinorial derivatives and using the definition of the components \re{comps} we find a supersymmetric
sigma model with $K$ as a potential for the metric:
\begin{align}\nonumber
&&S = \int d^2x \Bigg[\partial_{\+}\phi^{a} G_{ab}\partial_{=}\phi^{b}+i{\textstyle\frac 1 2}
(\psi_+^{a}\nabla_{=}\psi_+^{b}+\psi_-^{a}\nabla_{\+}\psi_-^{b})G_{ab}\\[1mm]\label{compact}
&&\qquad\qquad~~~~ -{\textstyle\frac 1 4} R_{cdab}\psi_+^{a}\psi_+^{b}\psi_-^{c}\psi_-^{d}\Bigg]
\end{align}
after eliminating the auxiliary fields ${\cal F}^a$. Here $a=(i, \bar i)$ etc. 
The geometry is a particular complex geometry called {\em K\"ahler} geometry. The metric, Levi-Civita connection and curvature tensor are all expressible in terms of the potential $K$:
\begin{equation}
G_{i~\!\bar j}=\partial^2 K/\partial\phi^i\partial\bar\phi^{\bar j}=:K,_{i\bar j}
\end{equation}
\begin{equation}
\Gamma_{ij}^{~~~k}=G^{k\bar s}\partial_{i}G_{j\bar s}=K^{k\bar s}K,_{ij\bar s}~,
\end{equation}
\begin{equation}
R_{i~\!\bar jk\bar s}=G_{m\bar j}\partial_{\bar s}(\Gamma_{ik}^{~~~m})
=K,_{i~\!\bar j k\bar s}-\Gamma_{ik}^{~~~m}\Gamma_{~~\bar j\bar s}^{\bar n}K,_{m\bar n}
\end{equation}
\smallskip
Here comma denotes derivative with respect to the fields indicated by the indices and $G^{i~\!\bar j}=K^{i~\!\bar j}$ is the inverse metric.
\begin{svgraybox}
K\"ahler geometry is the target space geometry of ${\cal N}=1$ sigma models in $4d$ and for chiral $ (2,2)$  sigma models in $2d$. The relation is $1-1$.
~
\end{svgraybox}
Here is a quick reminder of the definition of K\"ahler geometry:
A manifold carrying a  complex structure $J$ and a metric $g$ hermitean with respect to the complex structure
\begin{equation}
J^tgJ=g~,
\end{equation}
is called a K\"ahler manifold if the complex structure is annihilated by the Levi-Civita connection
\begin{equation}
\nabla J=0~.
\end{equation}
The metric then has a potential $K$ such that $g_{i\bar j}=K,_{i\bar j}$ in complex coordinates. Further there is a (globally defined) symplectic form
\begin{equation}
\omega=g J~,
\end{equation}
called the K\"ahler form.~\\

In \re{compact} the reduction is all the way from $(2,2)$ to components. But if we are only interested in the geometry as defined by the metric and $B$-field, it is already displayed in the reduction from $(2,2)$ to $(1,1)$ superspace.  This is generally true for our $(2,2)$ sigma models and now we demonstrate it for the chiral sigma model.

The $(1,1)$ superfields are $\Phi=\Phi_|$, where the vertical bar denotes setting half of the $(1,1)$ Fermi coordinates to zero $\theta-\bar\theta=0$. The spinorial derivatives reduce as
\begin{align}\nonumber
&\bbD{\pm}=D_\pm-iQ_\pm\\[1mm]
&\bbDB{\pm}=D_\pm+iQ_\pm\\[1mm]
&\implies\bbD{~}^2\bbDB{~}^2\sim D^2Q^2
\end{align}
where the $D$s are the $(1,1)$ derivatives and the $Q$s generate the non manifest second supersymmetries.
The action becomes\\
\begin{equation}
S=\int d^2x \bbD{}^2\bbDB{}^2K(\Phi,\bar\Phi)_|\to\int d^2x ~\!D{}^2Q^2K(\Phi,\bar\Phi)_|~.
\end{equation}
We evaluate the action using 
\begin{equation}\label{Dred}
Q_\pm\Phi^a=J^a_bD_\pm\Phi^b~,~~~J=\left(\begin{array}{cc}i\delta^{ i}_{ j}&0\cr
0&-i\delta^{\bar i}_{\bar j}\end{array}\right)
\end{equation} 
which follows from the reduction of the chirality constraints. 
\begin{align}\nonumber
&0=\bbDB{\pm}\Phi^i=(D_\pm+iQ_\pm)\Phi^i\\[1mm]\nonumber~,~~~&\Rightarrow Q_\pm\Phi^i=iD_\pm\Phi^i\\[1mm]
&\Rightarrow Q_\pm\bar\Phi^{\bar i}+=-iD_\pm\bar\Phi^{\bar i}
\end{align}
Here we have again assumed that there are $d$ chiral and their $d$ complex conjugate antichiral fields labeled by $\Phi^a=(\Phi^i,\bar \Phi^{\bar i})$.
Using this and integrating by parts we find the $(1,1)$ action:
\begin{svgraybox}
\begin{equation}
\int d^2x D{}^2Q^2K(\Phi,\bar\Phi)_|=\int d^2x D{}^2\Big(D_+\Phi^iK,_{i\bar j}(\Phi)D_-\Phi^{\bar j}\Big)~,
\end{equation}
\end{svgraybox}
\noindent
where again comma denotes derivative.
Comparing to \re{oneoneact} we see that the second derivative matrix $K,_{i\bar j}$ is the complex metric.\\

~\\
\centerline{\em Chiral and Twisted Chiral models}
~\\

We now turn to a sigma model based on both chiral superfields, $\bbDB{\pm}\Phi=0$, and twisted chiral ones,
~~$\bbDB{+}\chi=0~,~~\bbD{-}\chi=0$~, \cite{Gates:1984nk}.\\

The $(1,1)$ reduction \re{Dred} implies the following form for the chirality constraints:
\begin{equation}
Q_\pm \Phi^a=J^{a}_{b}D_\pm\Phi^{b}~,~~~Q_\pm\chi^{a'}=\pm J^{a'}_{b'}D_\pm\chi^{b'}~.
\end{equation}
Consider a sigma model with both types of fields
\begin{svgraybox}
\begin{align}\nonumber
&&\int d^2x D{}^2Q^2K(\Phi,\bar\Phi,\chi,\bar\chi)_|=:\int d^2x D{}^2Q^2K(\bbX{} )_|\\[1mm]\nonumber
&&=\int d^2x D{}^2\Big(D_+\bbX{A}(G_{AB}+B_{AB})(\bbX{} )D_-\bbX{B}\Big)\\[1mm]
&&
=\int d^2x D{}^2\Big(D_+\bbX{A}E_{AB}(\bbX{} )D_-\bbX{B}\Big)
\end{align}
\end{svgraybox}
\noindent
Where we introduced the notation $\bbX{A}=(\Phi^a,\chi^{a'})=(\Phi^i,\bar \Phi ^{\bar i},\chi^{i'},\bar\chi^{\bar i'})$, with the indices ranging over $i=1,\dots, d~, i'=1,\dots, d'$.  The middle and last line refers to \re{oneoneact}.  The metric and $B$-field are then given by
\begin{equation}
E_{AB}=\left(\begin{array}{cccc}
0&K,_{i\bar j}&K,_{i j'}&0\cr
K,_{\bar i j}&0&0&K,_{\bar i\bar j'}\cr
-K,_{i' j}&0&0&-K,_{i' \bar j'}\cr
0&-K,_{\bar i' \bar j}&-K,_{\bar i'  j'}&0\cr
\end{array}\right)
\end{equation}
which leads to the following field strength
\begin{align}\nonumber
& H_{i\bar jk'}= K,_{i\bar jk'}~,~~~~~~~~H_{i\bar jk'}= -K,_{i\bar j\bar k'}\\[1mm]
& H_{i'\bar j'k}= -K,_{i'\bar j'k}~,~~~H_{i'\bar j'\bar k}= K,_{i'\bar j'\bar k}~.
\end{align}
Note that, as in the previous example, the Lagrangian $K$ is a potential for the metric and now also for the $B$ field/torsion.

The two complex structures can be read off from the non manifest transformations of  $\Phi$ and $\chi$
according to (suppressing indices)
\begin{equation}
\delta_\pm\bbX{}=\epsilon_{(\pm)}^{~~\alpha}\mathbb{J}^{(\pm)}D_\alpha
\end{equation}
which leads to
\begin{equation}
\mathbb{J}^{(+)}=\left(\begin{array}{cc}
J&0\cr
0&J
\end{array}\right)~,~~~\mathbb{J}^{(-)}=\left(\begin{array}{cccc}
J&0\cr
0&-J
\end{array}\right)~,
\end{equation}
where again $J=diag (i,-i)$.
It is easy to see that they commute $[\mathbb{J}^{(+)},\mathbb{J}^{(-)}]=0$ and a bit more effort shows that
\begin{equation}
\nabla_{(+)}\mathbb{J}^{(+)}=0~,~~~\nabla_{(-)}\mathbb{J}^{(-)}=0~,
\end{equation}
where $\pm$ refers to $\pm {\textstyle\frac 1 2} HG^{-1}$ torsion. 
Finally, a local product structure is seen to be
\begin{equation}
\mathbb{K}=-\mathbb{J}^{(+)}\mathbb{J}^{(-)}=\left(\begin{array}{cc}
\bf{1}&0\cr
0&-\bf{1}\end{array}\right)~.
\end{equation}
These structures define bihermitian geometry  \cite{Gates:1984nk} which we now recapitulate:\\
A manifold endowed with two complex structures $J_+$ and $J_-$, a metric $g$ and an antisymmetric $B$-field $B$ carries a  {\em bihermitian geometry} if $g$ is hermitian with respect to both complex structures
\begin{equation}
J^t_\pm g J_\pm=g~,
\end{equation}
and the two complex structures are covariantly constant with respect to two connections with torsion
\begin{equation}
\nabla^{(+)}J_+=0~,~~~\nabla^{(-)}J_-=0~,
\end{equation}
where the torsionfull connections are
\begin{equation}
\nabla^{(+)}=\nabla^{(0)}+T~,~~~~\nabla^{(-)}=\nabla^{(0)}-T~,~~~~~
T^{~~k}_{ij}={\textstyle \frac 1 2}H_{ijn}g^{nk}~,
\end{equation}
and $\nabla^{(0)}$ is the Levi-Civita connection for $g$.

There are two distinct cases of this geometry depending on whether the two complex structures commute or not.

When the complex structures commute,
$
[J_+,J_-]=0~,
$
they define a third structure, a local product structure $\mathbb{K}$, by
\begin{equation}
\mathbb{K}:=-J_+J_-~,~~~\Rightarrow \mathbb{K}^2=1~,~~~
\end{equation}
This geometry is sometimes called a  {\em BILP geometry} (for bihermitian local product). 

\begin{svgraybox}{
BILP geometry is the target space geometry of $(2,2)$ sigma models with  $B$ field and commuting complex structures. It becomes manifest when the model is written in terms of chiral and twisted chiral  superfields.
It is a special case of Generalised K\"ahler geometry.}
\end{svgraybox}

There is a generalisation to $p$ left and $q$ right complex structures: {$(p,q)$ \em Hermitean geometry} \cite{Hull:1985jv}.
~\\

\centerline{Semichiral superfields }
~\\

Semichiral superfields \cite{Buscher:1987uw} obey only half the chirality constraints compared to chirals and twisted chirals.
This is mirrored in the reduction to  $(1,1)$ superfields in that in addition to the constraints 
 \begin{align}\nonumber
&&\bbDB{+}\ell=(D_++iQ_+)\ell=0~,~~~\Rightarrow Q_+L=JD_+L\\[1mm]
&&\bbDB{-}\mathfrak{r}=(D_-+iQ_-)\mathfrak{r}=0~,~~~\Rightarrow Q_-R=JD_-R~,
\end{align}
where $L=(\ell,\bar\ell)$ and $R=(\mathfrak{r},\bar{\mathfrak{r}})$, there arise two unconstrained spinorial $(1,1)$ fields:
\begin{equation}
Q_-L=\Psi_-~,~~~~~Q_+R=\Psi_+
\end{equation}
These are auxiliary fields. When they are integrated out of an action they become part of the complex structures:
\begin{equation}
Q_-L=\Psi_-(L,R)~,~~~~~Q_+R=\Psi_+(L,R)~.
\end{equation}

A general action is
\begin{equation}S=\int d^2x \bbD{}^2\bbDB{}^2K(L,R)_|\to\int d^2x D{}^2Q^2K(L,R)_|
\end{equation}
Pushing in the $Q$s and using the definitions of the $(1,1)$ components gives, after integrating out the auxiliary spinors,
\begin{svgraybox}
\begin{equation}
\int d^2x~\! D{}^2\Big(D_+\bbX{A} E_{AB}(\bbX{})D_-\bbX{B}\Big)
\end{equation}
\end{svgraybox}
~

~\\
where $\bbX{A}=(L,R)=(\ell^a, \bar\ell^{\bar a}, \mathfrak{r}^a, \bar {\mathfrak{r}}^{\bar a})$. To integrate out $\Psi_\pm$ the number of left must equal the number of right semichiral fields\footnote{The case of different number of left and right fields can be related to $\beta\gamma$-systems interacting with sigma-models and is treated in \cite{Lindstrom:2020oow}.}
 $a=1.\dots, d$.
Integrating out $\Psi_\pm$ then results in  {\em non-linear} relations \cite{Lindstrom:2005zr}, \cite{Bogaerts:1999jc}~.

The metric plus $B$ field are
\begin{align}\nonumber
&E_{LL}=[J,K_{LL}]K^{LR}JK_{RL}\\[1mm]\nonumber
&E_{LR}=JK_{LR}J+[J,K_{LL}]K^{LR}[J,K_{RR}]\\[1mm]\nonumber
&E_{RL}=-K_{RL}JK^{LR}JK_{RL}\\[1mm]
&E_{RR}=-K_{RL}JK^{LR}[J,K_{RR}]~.
\end{align}
The notation here is a bit stenographic. $K_{LR}$ is the matrix of second 
derivatives of $K$ with respect to left and right fields and $K^{RL}$ is its inverse. So the Lagrangian is now a potential for all the geometry, albeit a nonlinear one.
The complex structures involve the commutator $C_{LL}:=[J,K_{LL}]$  with $J$ the canonical complex 
structure in the left sector, et.c. The complex structures read
\begin{equation}
\mathbb{J}_+=\left(\begin{array}{cc}J&0\cr
K^{RL}C_{LL}&K^{RL}JK_{LR}
\end{array}\right)~,~~~\mathbb{J}_-=\left(\begin{array}{cc}K^{LR}JK_{RL}&K^{LR}C_{RR}\cr
0&J
\end{array}\right)
\end{equation}
and will not commute in general.
\bigskip

\begin{svgraybox}
This describes bihermitean geometry for the symplectic case  where the complex structures do not commute. The general case involves chiral, twisted chiral and semichiral fields. ~~~~~$K\to K\big(\Phi,\chi,L,R\big)$
\end{svgraybox}
\noindent
The general case then covers all the three cases previously discussed and correspondingly there is a geometry that includes all the hitherto discussed complex geometries. We now turn to that geometry.
~\\
~\\
\centerline{\em Generalised Complex Geometry}

~\\
In {\em generalised complex geometry}  \cite{Hitchin:2003cxu}, \cite{Gualtieri:2003dx}, the tangent bundle $T{\cal{M}}$ is replaced by the sum of the tangent and cotangent bundles,
\begin{equation}
{\mathbb{T}}:= T{\cal{M}}\oplus T^*{\cal{M}}~,
\end{equation}
 called the {\em generalised tangent bundle}\footnote{See Fig.1 below for the local structure around a regular point.}. Elements of $\mathbb{X}\in{\mathbb{T}}$ may be written as 
 \begin{equation}
 \mathbb{X}=X+\xi~,
 \end{equation}
 where 
 \begin{equation}
 X\in T{\cal{M}}~,~~~\xi\in T^*{\cal{M}}~.
 \end{equation}
 Alternatively, it is often useful to write $\mathbb{X}$ as a column vector
 \begin{equation}\label{column}
 \mathbb{X}=\left(\begin{array}{c}X\\ \xi\end{array}\right)
 \end{equation}

A {\em generalised almost complex structure} on ${\cal{M}}$ is an endomorphism ${\cal J}$ of the tangent bundle which squares to minus one 
 \begin{equation}
{{\cal J}} :~\mathbb{T}{\cal{M}}\longrightarrow \mathbb{T}{\cal{M}}~,~~~{\cal J}^2=-\bf{1}~,
\end{equation}
and preserves the natural pairing metric
\begin{align}\nonumber
&{\cal J}^t\eta {\cal J}=\eta~,~~~~\eta=\left(\begin{array}{cc} 0& \bf{1}\\ \bf{1}&0\end{array}\right)\\[2mm]
&\mathbb{Y}^t\eta \mathbb{X}=(Y, \sigma)\eta\left(\begin{array}{c}X\\ \xi\end{array}\right)
=Y^\mu\xi_\mu+\sigma_\mu X^\mu~.
\end{align}

The projection operators
\begin{equation}
\Pi_\pm:={\textstyle \frac 1 2}\left(\bf{1}\pm i{{\cal J}}\right)
\end{equation}
may be used to split the generalised tangent space in two parts at a point: 
The  $+i$  eigenspace $ {\mathbb {L}}$ and the  $-i$  eigenspace $\overline {\mathbb {L}}$
\begin{equation}
\mathbb {T}\otimes {\mathbb C} = {\mathbb {L}} \oplus \overline {\mathbb {L}}=\Pi{\mathbb {L}} \oplus\overline\Pi{\mathbb {L}}~.
\end{equation}
In complete analogy to
\begin{equation}
{T}\otimes {\mathbb C} = T^{(0,1)}\oplus T^{(1,0)}
\end{equation}
for the ordinary complex structure on $\cal{M}$.

The {\em Courant bracket} $\llbracket ~,~\rrbracket_C$ on ${\mathbb{T}}$ is defined by
\begin{equation}
\llbracket \mathbb{X},\mathbb{Y}\rrbracket_C=\llbracket X+\xi,Y+\eta\rrbracket_C=[X,Y]+{\cal L}_X\eta-{\cal L}_Y\xi-\frac{1}{2}d(i_X\eta-i_Y\xi),
\end{equation}
where $\mathbb{X},\mathbb{Y}\in C^\infty(\mathbb{T})$ and ${\cal L}_X$ is the Lie derivative with respect to $X$. 

The Courant bracket is antisymmetric in its arguments but does not satisfy the Jacobi identity.
The {\em Dorfman bracket}
\begin{equation}\label{dorf}
\llbracket \mathbb{X},\mathbb{Y}\rrbracket_D=[X,Y]+{\cal L}_X d\eta-i_{Y}d\xi~,
\end{equation}
does satisfy the Jacobi identity but is not antisymmetric. The relation between the brackets is 
\begin{equation}
\llbracket \mathbb{X},\mathbb{Y} \rrbracket_D=\llbracket \mathbb{X},\mathbb{Y} \rrbracket_C+d\eta( \mathbb{X},\mathbb{Y})~.
\end{equation}
Clearly, the brackets are equal when restricted to an isotropic subspace (i.e. a subspace ${\cal{M}}$ for which $\eta( \mathbb{X},\mathbb{Y})=0$ for all $\mathbb{X},\mathbb{Y}\in \cal{M}$). 

An important feature of the Courant bracket is that its automorphisms include $B$-transforms by closed two forms $B$.

\begin{equation}
\llbracket e^B\mathbb{X},e^B\mathbb{Y} \rrbracket_C=e^B\llbracket \mathbb{X},\mathbb{Y} \rrbracket_C+i_Yi_XdB~,
\end{equation}
where the last term vanishes precisely when $dB=0$.  In the representation \re{column} we have
\begin{equation}
e^B=\left(\begin{array}{cc} 1&0\\B&1\end{array}\right)
\end{equation}
This automorphism corresponds to the $B$-field gauge transformations in the sigma model \cite{Gualtieri:2014kja}.

Both the Courant and the Dorfman brackets may be twisted by a closed three form $H$:
\begin{equation}
\llbracket \mathbb{X},\mathbb{Y} \rrbracket
\to
\llbracket \mathbb{X},\mathbb{Y} \rrbracket+i_Yi_XH.
\end{equation}
Integrability of a generalised almost complex structure ${\cal J}$ is defined by requiring  that the subspaces defined by the projections are involutive\footnote{This definition of integrability runs parallel to  the usual one for a hermitian space $({\cal M},g,J)$ if we take $g\to\eta$ and $J\to{\cal J}$ and replace the Lie bracket by the Courant bracket.} i.e., that
\begin{equation}
\Pi_\mp\llbracket \Pi_\pm\mathbb{X},\Pi_\pm\mathbb{Y}\rrbracket_C=0~,~~~\forall ~ \mathbb{X},\mathbb{Y}\in C^\infty(\mathbb{T}).
\end{equation}

This requires the vanishing of the {\em generalised Nijenhuis  tensor} $\mathbb{N}$. In index free notation 
 \begin{equation}
{\mathbb{N}}_{{\cal J}}(\mathbb{X},\mathbb{Y})=\llbracket \mathbb{X},\mathbb{Y}\rrbracket_C+{\cal J}\llbracket {\cal J}\mathbb{X},\mathbb{Y}\rrbracket_C+{\cal J}\llbracket\mathbb{X}, {\cal J}\mathbb{Y}\rrbracket_C- \llbracket{\cal J}\mathbb{X}, {\cal J}\mathbb{Y}\rrbracket_C=0~.
\end{equation}
~\\
It is sometimes advantageous to define integrability with respect to the $H$-twisted bracket.
An integrable generalised almost complex structure is called a {\em generalised complex structure}.\\

Comment: A generalised complex structure ${\cal{J}}$ comes associated with a Poisson structure which sits in the upper right quadrangle when ${\cal{J}}$ is viewed as a matrix, mapping
\begin{equation}
{\mathbb{T}}\to {\mathbb{T}}~.
\end{equation}
 An {\em irregular point} is a point where this Poisson structure changes rank. A lot of the mathematical interest is focused on these points, but we will not comment on them further in this presentation.

~\\
A complex geometry $({\cal M}, J)$ or a symplectic geometry $({\cal M}, \omega)$ are  seen to be special cases of generalised geometry where the generalised complex structures are
\begin{equation}
{\cal J}_J=\left(\begin{array}{cc}J&0\\0&-J^t\end{array}\right)~,
\end{equation}
and 
\begin{equation}
{\cal J}_\omega=\left(\begin{array}{cc}0&-\omega^{-1}\\\omega&0\end{array}\right)~,
\end{equation}
respectively.
\begin{figure}[h]
\sidecaption
\includegraphics[scale=.65]{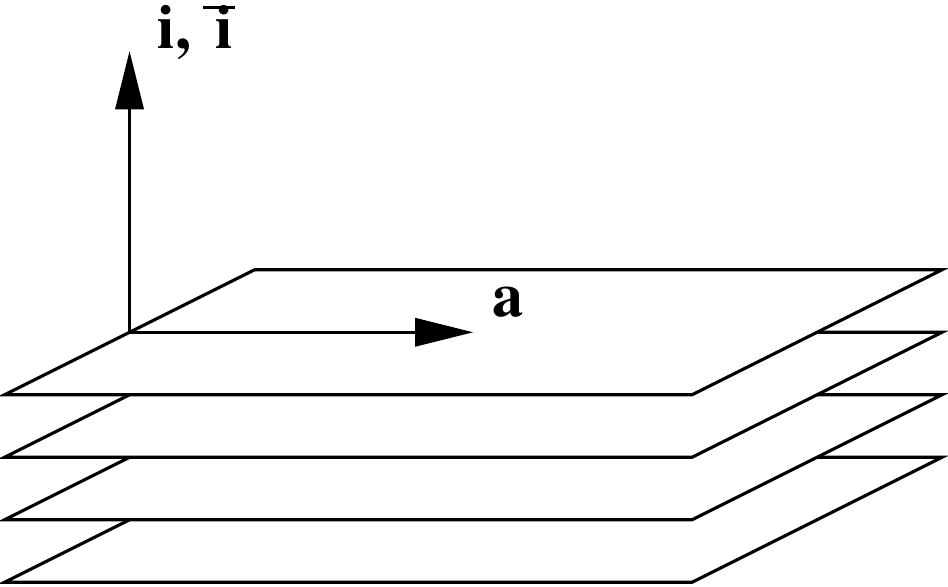}
%
%
\caption{A generalisation of the Newlander-Nirenberg theorem shows that locally, in the neighbourhood of a regular point. a GC manifold $\cal{M}$ looks like a foliation with $z^i,\bar z^{\bar i}$ complex and $x^a$ Darboux coordinates.}
\label{fig:1}       
\end{figure}
\bigskip

{\em Generalised K\"ahler Geometry} \cite{Gualtieri:2014kja}, is the form bihermitian geometry takes when lifted to generalised complex geometry. The additional data  is that there are {\em two} generalised complex structures, ${\cal J}_1$ and ${\cal J}_2$, that {\em commute}, $[{\cal J}_1,{\cal J}_2]=0, $ and whose product gives an integrable local product structure $\cal{G}$
\begin{equation}
{\cal G}=-{\cal J}_1{\cal J}_2~,~~~{\cal G}^2=\bf{1}~.
\end{equation}
Integrability is again defined with respect to the  Courant bracket or its H-twisted version. Under quite general conditions the existence on ${\cal M}$ of a metric $g$ acting on $T$, with inverse $g^{-1}$ that acts on $T^*$, ensures that one can find coordinates and $B$-transforms such that \cite{Gil}
\begin{equation}
{\cal G}=\left( \begin{array}{cc}0&g^{-1}\\
g&0\end{array}\right)~.
\end{equation}

Using ${\cal G}$ we may introduce  projection operators
\begin{equation}
P_\pm :={\textstyle \frac 1 2}\left(1\pm {\cal G}\right)~,
\end{equation}
that may be used to split the generalised tangent space $\mathbb{T}$ into the $\pm 1$ eigenspaces of ${\cal G}$:
\begin{equation}
\mathbb {T}=\mathbb {T}_+\oplus\mathbb {T}_-~,
\end{equation}
where
\begin{equation}
\mathbb {T}_\pm:=P_\pm\mathbb {T}~.
\end{equation}
~\\

\centerline{\em Gualtieri's map}
\bigskip

In \cite{Gualtieri:2014kja} the relation between bihermitian geometry $({\cal M}, G,J_\pm)$ and generalised K\"ahler geometry was established.
Explicitly the map constructs the two generalised complex structures ${\cal J}_1$ and ${\cal J}_2$ from the bihermitian data as follows:
\begin{equation}\label{gualtmap}
{\cal J}_{1/2}={\textstyle \frac 1 2}\left( \begin{array}{cc}J_+&-(\omega)_+^{-1}\\
\omega_+&-J_+^t \end{array}\right)
\pm{\textstyle \frac 1 2}\left( \begin{array}{cc}J_-&-(\omega)_-^{-1}\\
\omega_-&-J_-^t \end{array}\right)~,
\end{equation}
where $\omega_\pm=gJ_\pm$ are the two-forms associated with $J_\pm$.\\
~\\
Notice that the $B$ field only enters the definitions via $H$ in the integrability conditions  
\begin{equation}
\Pi_\mp^{1/2}\llbracket \Pi_\pm^{1/2}\mathbb{X},\Pi_\pm^{1/2}\mathbb{Y}\rrbracket_H=0~,~~~\forall ~ \mathbb{X},\mathbb{Y}\in C^\infty(\mathbb{T}).
\end{equation}
where
\begin{equation}
\Pi_\mp^{1/2}:={\textstyle \frac 1 2}\left(1\pm i{\cal J}^{1/2}\right)~.
\end{equation}
If integrability is defined with respect to the untwisted Courant bracket the relation \re{gualtmap} will involve $B$-transforms with a non exact $B$ instead.

\section{Concluding comments}

Generalised K\"ahler geometry covers the target space geometry of all three types of $(2,2)$ sigma  models described in the previous section, as well as the general one that includes all three types of fields. For different number of left $q$ and right $p$  supersymmetries, i.e., $(p,q)$ supersymmetry, the geometries will differ. Some cases are listed in table 1.
~\\

\centerline{{\bf Table 1}  Geometries of some $2d$ sigma-models with $(p,p)$ supersymmetries.}
\begin{table}[h]
\centering
\begin{tabular}{|l|c|c|c|c|c|}
\hline
Susy &(1,1) & (2,2) &{(2,2)}& (4,4) & (4,4) \\ 
\hline
E=G+B & $G,B$ &  $G$ & ${G,B}$ & $G$ & $G,B$ \\
\hline
Geometry   &  Rieman & K\"ahler & bihermitian  & hyperk\"ahler & bihypercomplex\\
\hline
\end{tabular}
~\\
\end{table}
Other cases such as Stong K\"ahler with torsion \cite{Howe:1988cj} or Strong hyperk\"ahler with torsion can also be described in generalised geometry \cite{Hull:2018jkr}, \cite{Gil}.

These relations between supersymmetric sigma models and geometry are utilised, e.g., in constructing quotients \cite{Lindstrom:1983rt}, \cite{Hitchin:1986ea} and $T$-dual models \cite{Rocek:1991ps},
\cite{Lindstrom:2007sq} .  Both these directions require the gauging of isometries of the models, studied in \cite{Hull:1985pq} \cite{deWit:2001bk} and, more relevant to generalised geometry, in \cite{Lindstrom:2007sq}, \cite{Merrell:2007sr}. Yet another issue concerns quantisation and is partly discussed in \cite{Grisaru:1997pg}, \cite{Grisaru:1997ep}.

\bigskip

\begin{acknowledgement}
I thank Martin Ro\v cek for reading and commenting on the manuscript. My research  is supported in part by the 2236 Co-Funded 
Scheme2 (CoCirculation2) of T\"UB{\.I}TAK (Project No:120C067)\footnote{\tiny However 
the entire responsibility for the publication is ours. The financial support received from 
T\"UB{\.I}TAK does not mean that the content of the publication is approved in a scientific 
sense by T\"UB{\.I}TAK.}. 
\end{acknowledgement}

\end{document}